\documentclass{article}

\usepackage[utf8]{inputenc}
\usepackage[T1]{fontenc}
\usepackage{amsmath, amssymb}
\usepackage[colorlinks=false, linkcolor=black, citecolor=black, urlcolor=black, hidelinks]{hyperref}
\usepackage{geometry}
\usepackage{graphicx}
\usepackage{authblk}
\usepackage[numbers,square]{natbib}
\usepackage{placeins}
\usepackage{algorithm, algorithmicx}
\usepackage{algpseudocode}

\title{\textbf{The effect of grain boundary misorientation on hydrogen flux using a phase-field based diffusion and trapping model}\\
{\small Published in: \href{https://doi.org/10.1002/adem.202401561}{Advanced Engineering Materials 26(22)(2024): 2401561}}}

\author[1]{Abdelrahman Hussein\thanks{Corresponding author: a.h.a.hussein@outlook.com}}
\author[2]{Byungki Kim}
\author[1]{Kim Verbeken}
\author[1]{Tom Depover}
\affil[1]{Department of Materials, Textiles and Chemical Engineering, Ghent University, Technologiepark 46, B-9052 Ghent, Belgium}
\affil[2]{School of Mechatronics Engineering, Korea University of Technology, Cheonan, Chungnam 31253, Republic of Korea}
\date{} 

\begin{document}

\maketitle

\begin{abstract}
  Understanding hydrogen-grain boundary (GB) interactions is critical to the analysis of hydrogen embrittlement in metals. This work presents a mesoscale fully kinetic model to investigate the effect of GB misorientation on hydrogen diffusion and trapping using phase-field based representative volume elements (RVEs). The flux equation consists of three terms: a diffusive term and two terms for high and low angle grain boundary (H/LAGB) trapping. Uptake simulations showed that decreasing the grain size resulted in higher hydrogen content due to increasing the GB density. Permeation simulations showed that GBs are high flux paths due to their higher enrichment with hydrogen. Since HAGBs have higher enrichment than LAGBs, due to their higher trap-binding energy, they generally have the highest hydrogen flux. Nevertheless, the flux shows a convoluted behavior as it depends on the local concentration, alignment of GB with external concentration gradient as well as the GB network connectivity. Finally, decreasing the grain size resulted in a larger break-through time and a larger steady-state exit flux.

\end{abstract}

\section{Introduction}

One of the challenges for realizing the full potential of hydrogen as a reliable clean energy carrier is hydrogen embrittlement (HE). This is a broad family of damage micromechanisms, which are activated by the accumulation (trapping) of hydrogen at different microstructure features, predominantly at grain boundaries (GBs) and interfaces \cite{Martin2012,Tunes2024, HiewSzeKei2023}. Yet, GBs and interfaces are complex and multifaceted features that can be categorized based on the type of the phases comprising the interface, their orientations, topology etc. 

These interfaces are critical hydrogen assisted crack initiation sites in several alloys. Triple junctions (TJs) were reported as crack initiation sites \cite{Koyama2014} in austenitic steels. Interaction of strain localization and interface decohesion were reported in dual-phase \cite{Koyama2014a}, ferritic \cite{Connolly2019} and pearlitic steels \cite{Li2024}. High angle grain boundaries (HAGBs) were found to be more susceptible to hydrogen assisted crack initiation and propagation compared to low angle grain boundaries (LAGBs) in Ni alloy 725 \cite{Taji2022}. Increasing the fraction of special low-energy grain boundaries improved the resistance to intergranular cracking in pure Ni \cite{Bechtle2009}. Hence, grain boundary engineering (GBE) is one of the primary strategies to alleviate the severity of HE \cite{Okada2023, Sun2023}. From this brief overview, a comprehensive GBE strategy requires a solid understanding of the two-way hydrogen GB interactions, i.e. how GBs retain hydrogen and how hydrogen initiates damage at the GBs. The former will be the topic of this article.

Traditionally, the kinetics of hydrogen diffusion and trapping are investigated using permeation experiments \cite{Eeckhout2023}. More novel techniques like atom probe tomography, scanning kelvin probe force microscopy and silver decoration could provide more spatial and temporal resolution of hydrogen-microstructure interactions \cite{Chen2020, Koyama2017a}. To understand the effect of GB misorientation on hydrogen flux, Koyama et al \cite{Koyama2017} used permeation experiments together with \emph{in-situ} silver decoration to spatially and kinetically resolve hydrogen mapping at the exit side of pure Fe. They found that GBs, especially HAGBs, are high flux paths compared to the lattice. This flux depends on the local diffusivity and concentration gradient. Since a constant diffusivity was assumed in pure Fe \cite{Hagi1979}, the increased GB hydrogen enrichment due to trapping resulted in such high flux. In order to have more quantitative understanding of such complex interactions, a fully kinetic mesoscale hydrogen diffusion and trapping model, which can resolve features like GBs and their type is required.

The most widely used models in the analysis of hydrogen transport are based on the macroscopic models of McNabb and Foster \cite{nabb1963}, Oriani \cite{Oriani1970}, which is coupled with mechanical effects by Sofronis and McMeeking \cite{Sofronis1989} and further modified by Krom et al. \cite{Krom1999}. Mesoscale models based on representative volume elements (RVEs) \cite{Diaz2019, Diaz2020, Hussein2023} can provide more quantitative understanding of such complex interactions.

To this end, a phase-field\footnote{It should be noted that in this work, the phase-field method is used to develop RVEs and represent microstructure features like grains and GBs, which is different than the phase-field fracture used in \cite{MartinezPaneda2018}.} based fully kinetic hydrogen diffusion and trapping model that can resolve the effect of HAGBs/LAGBs on hydrogen flux is presented in this work. First, starting from a total free energy functional in section \ref{Section: Model}, a fully kinetic hydrogen diffusion and trapping equation is derived. The computational setup together with the RVE generation and assigning orientations to account for GB misorientation is described in section \ref{Section: CompSetup}. Three RVEs were generated to study the effect of grain size. Additionally, two types of simulations were performed: uptake and permeation. The results of these simulations are discussed in section \ref{Section: Results} and were in good qualitative agreement with the experiments in \cite{Koyama2017}. The methodology described in this work lays the foundation for \emph{in-silico} framework for grain boundary engineered hydrogen resistant microstructures. 

\section{Modelling framework}
\label{Section: Model}

\subsection{Equilibrium and transport equations}

The phase-field method represents the grains by a vector of order parameters $\phi_i$, which take the values $0 \le \phi \le 1$. The principle of the phase-field based hydrogen diffusion and trapping model is to use functions of the phase-field order parameters as potentials, where their functional gradient induces a trapping flux. A fully kinetic equation for hydrogen diffusion and trapping at HAGB/LAGB can be then derived from the total free energy functional

\begin{equation}
  F = \int_V \Big[f^\mathrm{ch} + f^\mathrm{HAGB} + f^\mathrm{LAGB}\Big] dV
  \label{Eq: Functional}
\end{equation}

\noindent Where $f^\mathrm{ch}$ is the chemical free energy. The solvent atoms are considered fixed in their lattice position, and thus, do not contribute to the entnropy of mixing part of $f^\mathrm{ch}$. Using the compound energy formalism for a two-sublattice model, where the solvent atoms are assumed fixed in their lattice position, and considering only the ideal chemical free-energy

\begin{equation}
  f^\mathrm{ch} = (1 - \theta) G_{X:Va} + \theta G_{X:H} + \frac{NRT}{V_\mathrm{m}} \left[(1 - \theta) \ln(1-\theta) + \theta \ln \theta\right]
  \label{Eq: fchem}
\end{equation}

\noindent Where the solvent atom is $X$, $\theta$ is the hydrogen occupancy, hydrogen atoms (H) and vacancies (Va) occupy the interstitial sites, $N$ is the number of interstitial sites per lattice unit cell per solvent atom. For the case of the BCC lattice with tetrahedral interstitial sites $N=6$, while for FCC lattice with octahedral sites $N=4$ \cite{Krom2000}. $R$ is the universal gas constant, $T$ is the temperature and $V_\mathrm{m}$ is the molar volume of the host lattice. A double-obstacle potential can be used with the misorientation angle $\Delta \psi_{\alpha \beta}$ between two grains $\alpha$ and $\beta$ for calculating the HAGBs function $g(\phi)_\mathrm{HAGB}$ and LAGBs function $g(\phi)_\mathrm{LAGB}$ according to \textbf{Algorithm} \ref{Alg: gPhi}. $f^\mathrm{HAGB}$ is the HAGB trapping energy density, which is constructed to decrease with increasing hydrogen occupancy

\begin{equation}
  f^\mathrm{HAGB} = g(\phi)_\mathrm{HAGB}(1 - \kappa_\mathrm{HAGB} \theta)
\end{equation}

\noindent Where $\phi$ is the vector of phase-field variables representing each grain, which takes the values $0 \le \phi \le 1$. $g(\phi)_\mathrm{HAGB}$ is the double-obstacle potential representing the HAGB. $\kappa_\mathrm{HAGB}$ is a parameter related to the HAGB trap-binding energy that will be evaluated from equilibrium as described later. Similarly, $f^\mathrm{LAGB}$ is  the LAGB trapping energy density, which is expressed as 

\begin{equation}
  f^\mathrm{LAGB} = g(\phi)_\mathrm{LAGB}(1 - \kappa_\mathrm{LAGB} \theta)
\end{equation}

\begin{algorithm}
  \caption{Double-obstacle potential for HAGB/LAGB.}\label{Alg: gPhi}
\begin{algorithmic}
  \For{$\alpha \in \{\phi_1, \dots, \phi_{N-1}\}$}
    \For{$\beta \in \{\phi_{\alpha+1}, \dots, \phi_N\}$}
      \If{$\Delta \psi_{\alpha \beta} \ge 15^\circ$} 
        \State $g(\phi)_\mathrm{HAGB} \mathrel{+}= \phi_\alpha \phi_\beta$
      \Else
      \State $g(\phi)_\mathrm{LAGB} \mathrel{+}= \phi_\alpha \phi_\beta$
      \EndIf
    \EndFor
  \EndFor
\end{algorithmic}
\end{algorithm}

\noindent Although this study is limited to HAGB/LAGB, the previous algorithm can be modified to account for more complicated interfaces, for instance coincident site lattice boundaries and/or interfaces between two different phases. 

The equilibrium is established by a multi-step parallel tangents between the chemical potentials of the lattice and each trapping site separately. First, a parallel tangent between the chemical potential of the lattice $\mu_\mathrm{L}$ far away from the GBs and that at the center of the HAGB, $\mu_\mathrm{HAGB}$, where $g(\phi)_\mathrm{HAGB}$ = 0 and 1/4, respectively, results in 

\begin{equation}
  \frac{\theta^\mathrm{center}_\mathrm{HAGB}}{1-\theta^\mathrm{center}_\mathrm{HAGB}} = \frac{\theta_\mathrm{L}}{1-\theta_\mathrm{L}} \exp\Big(\frac{\kappa_\mathrm{HAGB} V_\mathrm{m}}{4NRT}\Big)
  \label{Eq: EquilHAGB_Center}
\end{equation}

\noindent And thus, $\kappa_\mathrm{HAGB}$ is related to the HAGB trap-binding energy $\Delta E_\mathrm{HAGB}$ as

\begin{equation}
  \kappa_\mathrm{HAGB} = -\frac{4N \Delta E_\mathrm{HAGB}}{V_\mathrm{m}}
  \label{Eq: kappa_HAGB}
\end{equation}

\noindent Similarly, for LAGBs

\begin{equation}
  \frac{\theta^\mathrm{center}_\mathrm{LAGB}}{1-\theta^\mathrm{center}_\mathrm{LAGB}} = \frac{\theta_\mathrm{L}}{1-\theta_\mathrm{L}} \exp\Big(\frac{\kappa_\mathrm{LAGB} V_\mathrm{m}}{4NRT}\Big)
  \label{Eq: EquilLAGB_Center}
\end{equation}

\noindent And 
\begin{equation}
  \kappa_\mathrm{LAGB} = -\frac{4N \Delta E_\mathrm{LAGB}}{V_\mathrm{m}}
  \label{Eq: kappa_LAGB}
\end{equation}

\noindent Using Equation \ref{Eq: EquilHAGB_Center}-\ref{Eq: kappa_LAGB}, the multi-trap equilibrium occupancy $\theta^\mathrm{eq}$ over all the RVE is

\begin{equation}
  \frac{\theta^\mathrm{eq}}{1-\theta^\mathrm{eq}} = \frac{\theta^\mathrm{eq}_\mathrm{L}}{1-\theta^\mathrm{eq}_\mathrm{L}} \exp\Big(\frac{\kappa_\mathrm{HAGB} V_\mathrm{m} g(\phi)_\mathrm{HAGB}}{NRT}
   + \frac{\kappa_\mathrm{LAGB} V_\mathrm{m} g(\phi)_\mathrm{LAGB}}{NRT} \Big)
  \label{Eq: EquilGB}
\end{equation}

From linear irreversible thermodynamics, the flux is given by  

\begin{equation}
  \vec{J} = -\theta M \nabla \Big( \frac{\delta F}{\delta \theta} \Big) 
\end{equation}

\noindent Where $M$ is the diffusion mobility. In the dilute limit, the diffusivity is related to the mobility by $M=D/RT$, consequently

\begin{equation}
  \vec{J} = -  D \Big( \frac{N}{V_\mathrm{m}} \nabla \theta - \frac{\theta \kappa_\mathrm{HAGB}}{RT} \nabla g(\phi)_\mathrm{HAGB} 
  - \frac{\theta \kappa_\mathrm{LAGB}}{RT} \nabla g(\phi)_\mathrm{LAGB}\Big) 
  \label{Eq: flux}
\end{equation}

\noindent Therefore, the total flux consists of three interacting flux terms 

\begin{equation}
  \vec{J} = \vec{J}_\theta +\vec{J}_\mathrm{HAGB} +\vec{J}_\mathrm{LAGB}
  \label{Eq: FluxTerms}
\end{equation}

\noindent Where $\vec{J}_\theta$ is the diffusive flux term with occupancy gradient as the driving force, which is always directed from high to low occupancy. The other two flux terms are the HAGB and LAGB trapping terms, respectively, with the double-obstacle potential gradients as the driving forces, and are always directed towards the center of the GB. Finally, the time evolution of the occupancy is obtained from

\begin{equation}
  \frac{\partial \theta}{\partial t} = -\nabla \cdot \vec{J} 
  \label{Eq: diffusion}
\end{equation}

\subsection{Suggested correction for numerical interface thickness}
\label{Section: Correction}

\begin{figure}[hbt!]
  \includegraphics[width=8cm]{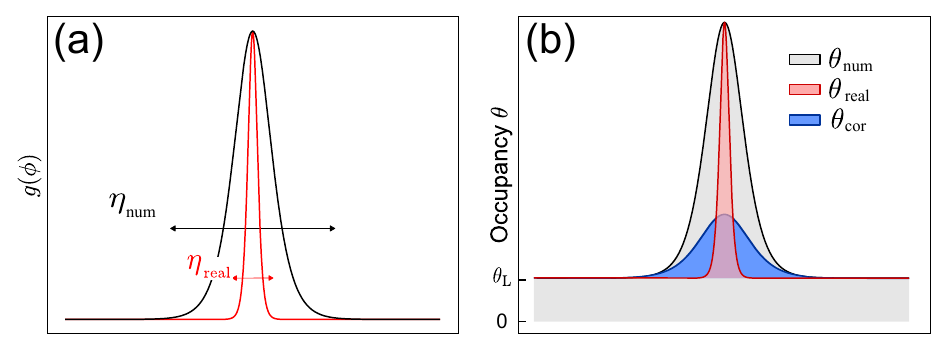}
  \centering
  \caption{Schematic showing (a) the effect of the numerical and real interface thickness (not to scale) and (b) the corrected occupancy profile $\theta_\mathrm{cor}$.}
  \label{Fig: Correction}
\end{figure}

One of the characteristics of the phase-field method is the diffuse interface concept for numerical convenience, i.e. using much larger values for the numerical interface width $\eta$, which can be in the order of several micrometers compared to the real interface width $\eta_\mathrm{real}$, which can be in the order of angstroms, as shown schematically in Fig.(\ref{Fig: Correction}a). This will lead to an overestimation of the occupancy represented by the area under the curve $\int^{\eta_\mathrm{num}}\theta dx= \theta_\mathrm{num}$, compared to the real occupancy $\int^{\eta_\mathrm{real}}\theta dx= \theta_\mathrm{real}$. Although numerical values can be used with the understanding that this is an overestimation and can provide very useful semi-quantitative information on hydrogen diffusion and trapping, accounting for this overestimation could provide improved outcome.

To remediate this shortcoming, while benefiting from the numerical convenience and the length-scale of the phase-field method at the mesoscale, a correcting scaling factor ${\lambda}$ can be used, such that

\begin{equation}
  \kappa^\mathrm{cor}_\mathrm{gb} = \lambda \kappa_\mathrm{gb}
  \label{Eq: kappaCorrected}
\end{equation}

\noindent $\lambda$ should keep the area $\theta_\mathrm{cor}=\theta_\mathrm{real}$ while in the same time use the numerical GB thickness $\eta = \eta_\mathrm{num}$ as shown in Fig.(\ref{Fig: Correction}). However, this will come at the cost of reducing the maximum peak of the occupancy, i.e. using a scaled $\Delta E_\mathrm{gb}$. Kim et al \cite{Kim2008} used the factor $\lambda = \eta_\mathrm{real}/\eta_\mathrm{num}$. Kundin et al. \cite{Kundin2021} used the factor $\lambda = \eta_\mathrm{real}/8 \eta_\mathrm{num}$. However, in our calculations these expressions did not result in the equal area condition $\theta_\mathrm{cor}=\theta_\mathrm{real}$. This is because they were used for substitutional solid solutions with $\Delta E_\mathrm{gb}$ much smaller than that for the interstitial hydrogen. Our empirical calculations showed that $\lambda =1/3$ provides satisfactory results in terms of the equal area condition for the spectrum of the GB trap-binding energies in the range -10 $-$ -60 kJ/mol and the value for $\eta_\mathrm{num}=5\mu$m that is use in our simulations as will be described next. This value for $\lambda$ will be used henceforth.

\section{Computational setup}
\label{Section: CompSetup}

The 2D RVE is generated using the multiphase-field method \cite{Steinbach2009} for normal grain growth according to the free-energy functional

\begin{equation}
  F = \int_V \Bigl[\sum_{\alpha=1}\sum_{\beta>\alpha} \frac{4 \sigma_{\alpha \beta}}{\eta} \Bigl(-\frac{\eta^2}{\pi^2} \nabla\phi_\alpha \cdot \nabla\phi_\beta + \phi_\alpha \phi_\beta \Bigr) \Bigr] dV
  \label{equation: Functional2}
\end{equation}

\noindent Where $\sigma_{\alpha \beta}$ is the interface energy between phases $\alpha$ and $\beta$. The evolution of $\phi$ is given by 

\begin{equation}
  \dot{\phi} = \frac{\pi^2}{8 \eta} \sum^N_{\alpha \neq \beta} \frac{\mu_{\alpha \beta}}{N} \left( \frac{\delta F}{\delta \phi_\alpha} - \frac{\delta F}{\delta \phi_\beta} \right)
\label{Eq:Evolution}
\end{equation}

\noindent Where $\mu_{\alpha \beta}$ is the interface mobility. The interface energy and interface mobility were 0.5 $\mathrm{Jm^{-2}}$ and $10^{-14} \ \mathrm{m^{2}J^{-1}s^{-1}}$, respectively. The numerical interface thickness $\eta = \eta_\mathrm{num}=5 \,\mu$m as discussed in section \ref{Section: Correction}

\begin{figure}[hbt!]
  \includegraphics[width=14cm]{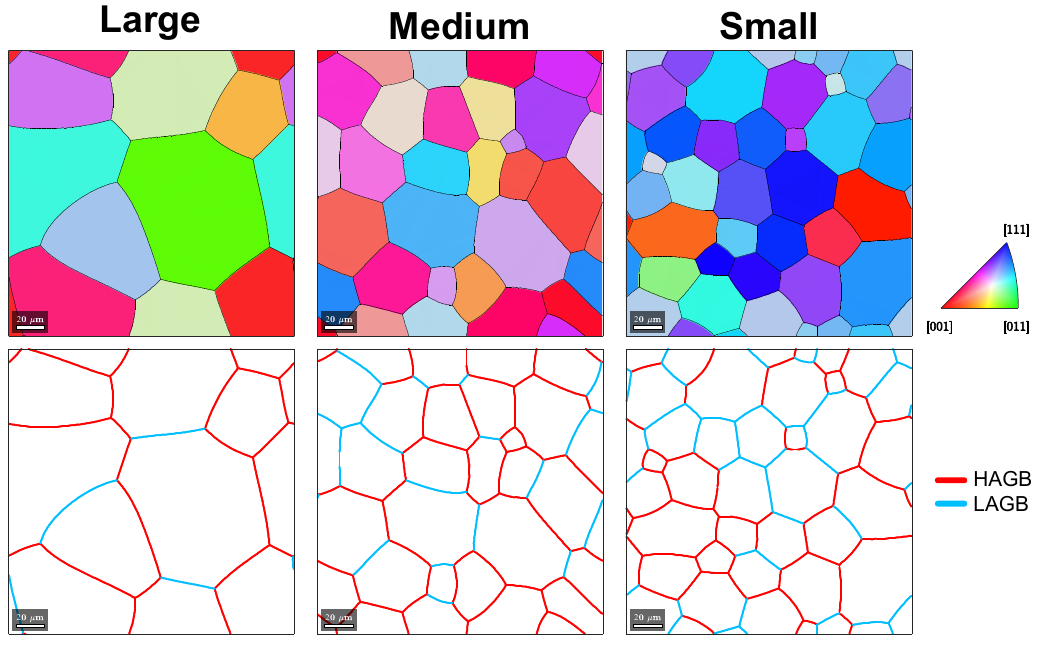}
  \centering
  \caption{The generated RVEs with three different grain sizes. The first row shows the inverse pole figure maps and the second row shows the GB misorientation map, where the HAGBs with misorientations $\Delta \psi \ge 15^\circ$ are shown in red color, while the LAGBs with $\Delta \psi < 15^\circ$ are represented by sky blue color.} 
  \label{Fig: RVE_mori}
\end{figure}

Three RVEs were generated with different grain sizes referred to as Large, Medium and Small as shown in \textbf{Figure} \ref{Fig: RVE_mori}. To model the effect of misorientations, MTEX \cite{Bachmann2010} is used to create orientation distribution functions (ODFs). Discrete orientations were then sampled from these ODFs and assigned to each grain. The inverse pole figure maps of the three RVEs with the classification of GBs to HAGB/LAGB is shown in Figure \ref{Fig: RVE_mori}. Sato and Takai \cite{Sato2023} showed that the fraction of HAGBs increases with increasing the grain size, where LAGBs were annihilated with increasing the annealing temperature. The ODFs in this work were carefully chosen to reflect this trend. The fraction of HAGBs are 0.81, 0.76 and 0.67 for Large, Medium and Small RVEs, respectively.

An in-house software developed by the authors called phase-FIeld MATerials Simulator ($\phi$MATS) was used to run the simulations. A python interface is used for pre/post processing. The input data are then saved in Hierarchical Data Format version 5 (HDF5) \cite{hdf5} file. A C++ interface reads these data and uses the Portable, Extensible Toolkit for Scientific Computation (PETSc) \cite{petsc-web-page} as a partial differential equation solver. At the current stage, finite-difference method with 21-point stencil and iterative Generalized Minimal RESidual method (GMRES) solver. A central difference implicit backward Euler scheme is used for time integration. The advantage is that the time increment is unconditionally stable, which is computationally more efficient than explicit solvers. The time step results are written in the hdf5 file, which can then be post-processed using python and/or Paraview \cite{AHRENS2005}.

The simulations were performed according to \textbf{Equation} \ref{Eq: flux} and \ref{Eq: diffusion} on a 1000 $\times$ 1000 grid. For uptake simulations, a constant occupancy $4.08\times10^{-9}$, which is equivalent to lattice concentration $c_\mathrm{L} = 3.453\times10^{-3} \ \mathrm{mol/m^3} = 2.08\times10^{21}\mathrm{ \ atoms/m^3}$ \cite{Sofronis1989}, was applied to all edges of the RVE. For permeation simulations, it was applied to the left edge, while, zero occupancy was applied to the right edge. The upper and lower edges were subjected to zero flux boundary conditions. The parameters used for the simulations are shown in \textbf{Table} \ref{Table: 1D}.

\begin{table}[hbt!]
  \centering
  \caption{Parameters used in the simulations}
  \begin{tabular}{ll}

  \hline
  Parameter                       & Value                          \\ 
  \hline
  Lattice diffusivity $D_\mathrm{L}$ $\mathrm{(m^2/s)}$ & $1\times10^{-8}$    \\
  HAGB activation energy (kJ/mol) &  -43.7 \cite{Sato2023}\\
  LAGB activation energy (kJ/mol) &   -26 \cite{Pressouyre1979}\\
  Lattice molar volume $V_\mathrm{m}$ $\mathrm{(m^3/mol)}$ & $7.09\times10^{-6}$    \\
  Number of occupancy sites per lattice atom $N$  & 6 \\

  Temperature $T$   $\mathrm{(K)}$                  & 300      \\
  Universal gas constant $R$   $\mathrm{(J/mol K)}$      & 8.31 \\
  \hline               
  \end{tabular}
  \label{Table: 1D}
\end{table}


\section{Results and discussion}
\label{Section: Results}

\subsection{Uptake simulations}

\begin{figure}[hbt!]
  \includegraphics[width=14cm]{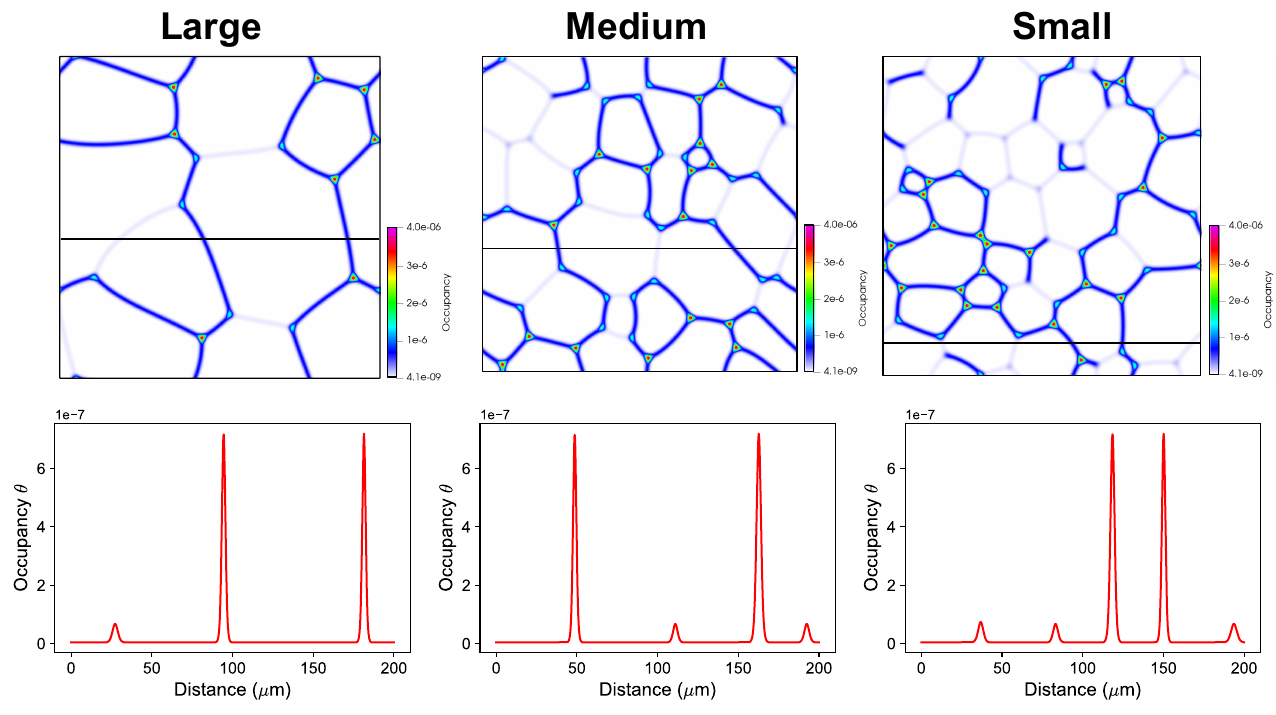}
  \centering
  \caption{Occupancy fields for the three RVEs for the uptake simulations are shown in the upper row. Line plots of occupancy profiles corresponding to the location of the black horizontal lines are in the second row.}
  \label{Fig: Uptake}
\end{figure}

The equilibrium occupancy fields, i.e. after reaching steady-state are shown in the first row of \textbf{Figure} \ref{Fig: Uptake}. It could be seen that HAGBs have higher occupancy than LAGBs. The line profiles in the second row of Figure \ref{Fig: Uptake} highlight this contrast. The base-line of these profiles is the lattice occupancy $\theta_\mathrm{L}$ and is equal to the boundary occupancy. At the triple junctions (TJs), $g(\phi) = 1/3$, while that at the GBs is 1/4. Substituting these values in \textbf{Equation} \ref{Eq: EquilGB} will result in higher occupancy. This is a useful outcome of the phase-field based RVEs, which can implicitly account for the larger trapping of TJs \cite{Zhou2021, Stender2011} without the need for the extra processing required to identify them in the geometric based RVEs \cite{Jothi2015}. Additionally, since there are two types of GBs; HAGBs and LAGBs, TJs will have different possible characters \cite{Fortier1997}. There are four types of TJs in the current RVEs, the ones between only HAGBs will have the largest occupancy. These will be followed by TJs composed of a pair of HAGBs and one LAGB, then those composed of a pair of LAGBs and one HAGB. The lowest TJ occupancy is when all the three GBs are LAGBs. This is an additional advantage of the phase-field approach, not only capturing TJs, but also their character. 

\begin{figure}[hbt!]
  \includegraphics[width=8cm]{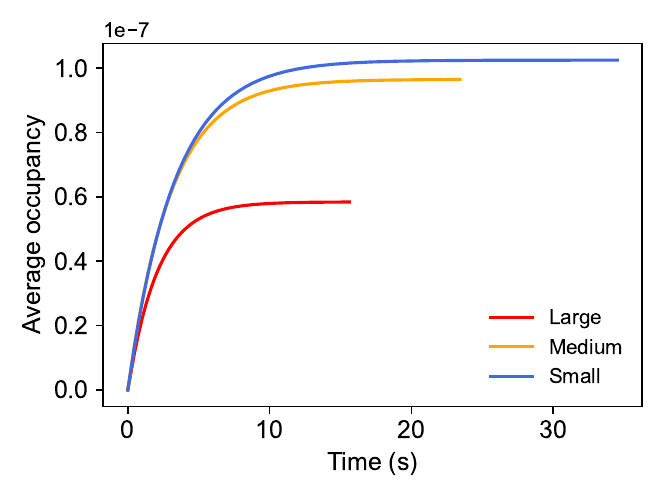}
  \centering
  \caption{Effect of grain size on the average hydrogen uptake.}
  \label{Fig: UptakeAverage}
\end{figure}

The time evolution of the average uptake occupancy $\langle\theta\rangle = 1/A \int \theta dA$, where $A$ is the area of the RVE, is shown in \textbf{Figure} \ref{Fig: UptakeAverage}. The average occupancy increases with decreasing the grain size. Furthermore, the time to reach equilibrium increases with decreasing grain size. This is because the GB density, i.e. trap density, increases with decreasing grain size, and thus, the total hydrogen uptake will increase \cite{Park2017}. It should be also noted that as mentioned earlier, the fraction of HAGBs decreases with decreasing grain size following the experimental observations reported in \cite{Sato2023}, however, these particular fractions do not significantly affect that trend. 

\subsection{Permeation simulations}

\begin{figure}[hbt!]
  \includegraphics[width=14cm]{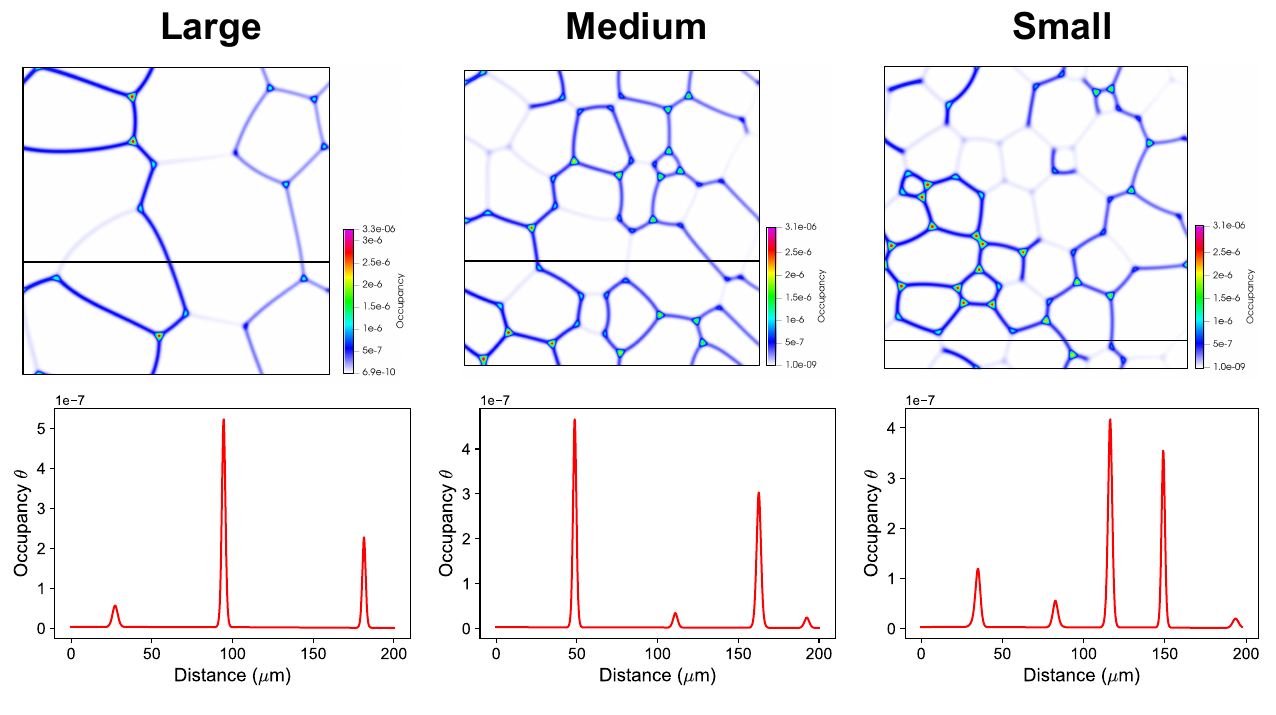}
  \centering
  \caption{Occupancy fields for the three RVEs after reaching steady-state in permeation simulations shown in the upper row. Line plots of occupancy profiles corresponding to the location of the black horizontal lines are in the second row. In both cases, the permeation is from left to right.}
  \label{Fig: Permeation}
\end{figure}

The steady-state occupancy distribution is shown in \textbf{Figure} \ref{Fig: Permeation}. It can be seen that the occupancy of the GBs near the entry side is larger than those closer to the exit side, as would be expected from a permeation setup. This is highlighted in the line plots at the second row of Figure \ref{Fig: Permeation}. It should be noted that the baselines of these profiles have a slope from the entry side (left) to the exit side (right) of $2.04 \times 10^{-11} \mu\mathrm{m^{-1}}$, however, it is not apparent since the maximum occupancy is four orders of magnitude higher.

\begin{figure}[hbt!]
  \includegraphics[width=16cm]{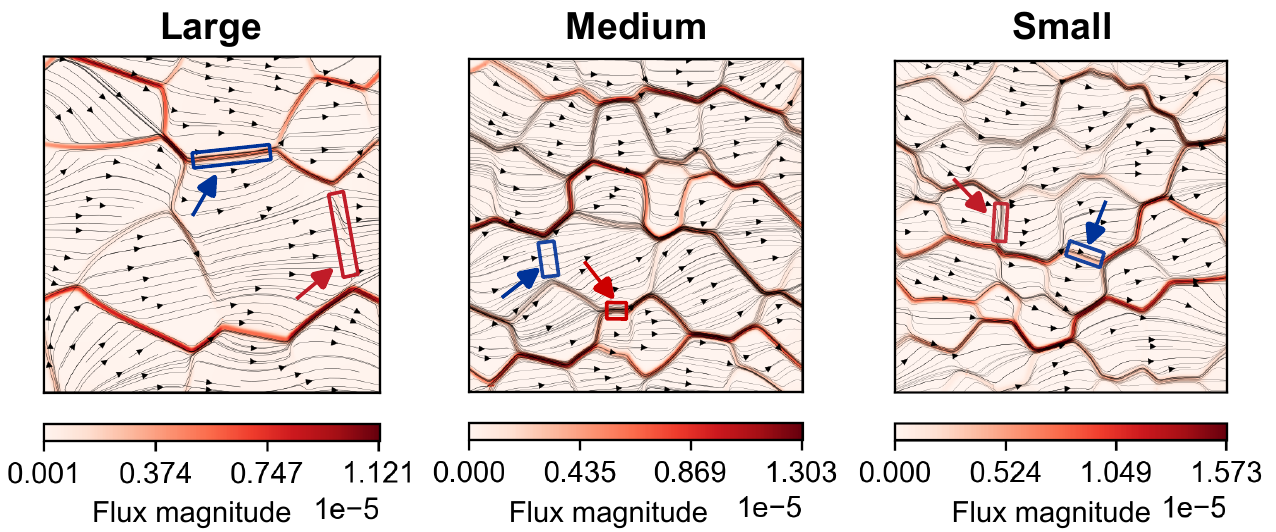}
  \centering
  \caption{Flux magnitude $|J|$ overlaid by the flux streamlines to represent the path of hydrogen within the RVEs. The white rectangles highlight LAGBs with relatively high $|J|$ due to their alignment and connectivity with two HAGBs. The red rectangles highlight HAGB with relatively low $|J|$ because they are perpendicular to the external occupancy gradient direction. }
  \label{Fig: Flux}
\end{figure}

The flux magnitude fields are shown in \textbf{Figure} \ref{Fig: Flux}. The streamlines of the flux are plotted on top of the magnitude of the flux fields. Streamlines are essentially flow paths which consist  of curves that are tangential to the flux vector at each point \cite{Jun2011}. It could be seen that, GBs have larger flux magnitude than the lattice. This is due to interaction of both the GB trapping and diffusion flux terms. The former leads to the accumulation of hydrogen at the GBs. This accumulation increases the occupancy gradient along the GBs, resulting in a large diffusive flux along the GBs. When there is no external occupancy gradient, i.e., the steady-state of uptake simulation, these two fluxes will cancel out each other. On the other hand, when there is an external occupancy gradient, as in the case of permeation simulation, the diffusive flux $\vec{J}_\theta$ will have slightly larger component along the GB compared to the trapping fluxes $\vec{J}_\mathrm{HAGB}$ and $\vec{J}_\mathrm{LAGB}$, resulting in a total flux $\vec{J}$ along the GB that is larger than that in the lattice. 

Since in this work there are two types of GBs, the HAGBs will accumulate more hydrogen than the LAGBs due to their larger trap-binding energy, resulting in a larger diffusive flux $\vec{J}_\theta$, and thus, overall total flux. This is in agreement with the observations and conclusions of Koyama et al. \cite{Koyama2017}. Following their same reasoning; that is, to rule out the effect of GB diffusivity on the total flux, a constant diffusivity is used according to the experimental observations in \cite{Hagi1979}. Thanks to the multi-term flux expression of \textbf{Equation} \ref{Eq: FluxTerms}, the complex interaction of HAGB/LAGB trapping and diffusive flux could be captured.

However, not only the trap-binding energy of the GB plays a role in determining the flux, but also geometric factors like the orientation of the GB with respect to the direction of the external occupancy gradient as well as the network connectivity of the GBs play an important role in determining the flux magnitude. The GBs that are more aligned with the external occupancy gradient have the largest flux, while the perpendicular ones have the lowest flux. Additionally, a more aligned GB path will generally have larger flux compared to a more tortuous path. For example, a LAGB that is aligned with the direction of the occupancy gradient and connected to two HAGBs that have approximately similar alignment, will have larger flux compared to a perpendicular HAGB, as marked by the blue and red rectangles, respectively in Figure \ref{Fig: Flux}.

\begin{figure}[hbt!]
  \includegraphics[width=15cm]{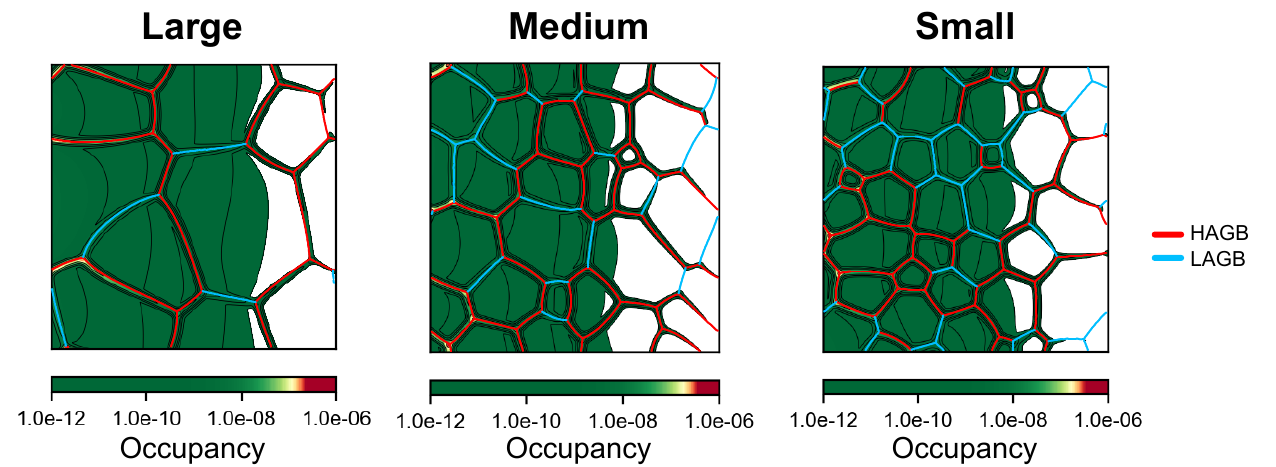}
  \centering
  \caption{Snapshots at early stages of the permeation simulation on a log-scale with a cut-off occupancy of $1\times10^{-12}$. The High/Low angle grain boundary maps are plotted on top of the occupancy to compare between the HAGB and LAGB paths.}
  \label{Fig: PermeationEarly}
\end{figure}

Additionally, based on the \emph{in-situ} silver decoration technique during permeation test, Koyama et al. \cite{Koyama2017} reported that at the exit side of the sample, silver particles deposited first on HAGBs, followed by LAGBs and finally the grain interior, i.e., the lattice. In other words, the hydrogen permeates faster along HAGBs followed by LAGBs then the lattice. In order to visualize this effect, snapshots of the occupancy were taken at the early stages of the permeation simulation as shown in \textbf{Figure} \ref{Fig: PermeationEarly}. To enhance visualization, the occupancy is plotted with a log-scale and a cut-off occupancy of $1\times10^{-12}$, thus, any occupancy lower than this value will have a white color. The black curves are contour lines of iso-occupancy to enhance visualizing the occupancy field. Additionally, the GB character is superimposed to highlight the GB type. Each snapshot was taken when an occupancy of $1\times10^{-12}$ reached the exit edge. A time-lapse of this early stage is shown in supplementary materials (S1-S3). 

As could be clearly seen in Figure \ref{Fig: PermeationEarly}, the order of permeation speed is in very good agreement with \cite{Koyama2017}. Since the Large RVE has only HAGBs at the exit edge, they cannot be compared to LAGBs, nevertheless, they are faster hydrogen paths than the lattice. Medium and Small RVEs have a combination of both HAGBs and LAGBs, it could be clearly seen that the order is HAGB$>$LAGB$>$lattice. 

The times for these snapshots were 0.195, 0.21 and 0.226 s for the Large, Medium and Small RVEs, respectively. Despite the relatively small differences, this means that hydrogen will take more time to permeate through the thickness with decreasing grain size. Although smaller grains have larger density of GBs, which are high flux paths as discussed, they are more tortuous, and thus, take more time to reach the exit side. Additionally, this time for the first hydrogen to reach the exit side is significantly smaller than the 30 min reported in \cite{Koyama2017}. The 200 $\mu$m thickness of the RVEs compared to the 340 $\mu$m sample thickness does not explain this large difference. This is likely due to the sensitivity of the silver decoration technique, which cannot detect the small flux of the early hydrogen permeation as reported in \cite{Koyama2017}. Indeed, they reported that the diffusion time is within 20 s, which is in good agreement with the time to reach steady-state flux as will be described next from the exit flux curves. 

\begin{figure}[hbt!]
  \includegraphics[width=8cm]{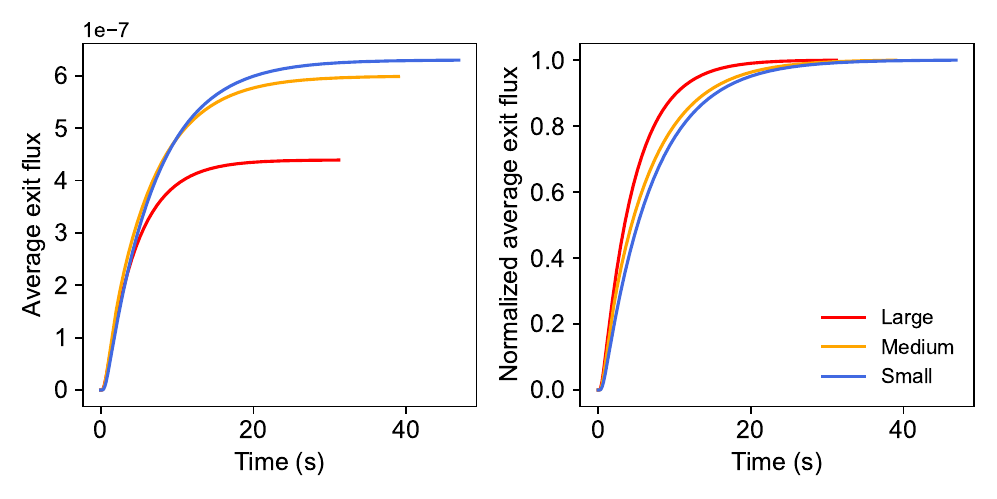}
  \centering
  \caption{Average and normalized exit flux curves for the permeation simulations.}
  \label{Fig: FluxAverageHLAGB}
\end{figure}

The average and the normalized exit flux at the exit side are shown in \textbf{Figure} \ref{Fig: FluxAverageHLAGB}. Reducing the grain size, which is equivalent to increasing the trap density, leads to increased steady-state flux. It also led to increased break-through time, which is defined as the time $t_{0.63}$ to reach 63\% of the steady-state flux, with values 19.69, 24.61 and 29.53 s for Large, Medium and Small RVEs, respectively. An increase in the break-through time was reported with increased trap-density for carbides \cite{Eeckhout2018} and dislocations \cite{Eeckhout2017}. However, since only the normalized flux was reported, it was not possible to correlate with the steady-state flux.

In terms of grain size in pure Fe, there were no permeation curves reported in the literature to the best of our knowledge, as also discussed in \cite{Diaz2019}. This is as opposed to the case for pure Ni \cite{Brass1996}, where it was reported that smaller grains have shorter break-through time and higher steady-state flux. The reason for the discrepancy with our results is that the GB diffusivity in Ni is higher than in the lattice, while the same diffusivity was used for both following \cite{Hagi1979}. Indeed, it was shown that increasing the GB diffusivity reduces the break-through time and increases the steady-state flux \cite{Hussein2023}. Since the density of the GBs increases with decreasing the grain size, longer time is needed for hydrogen to saturate the GBs, and thus, larger break-through time. Additionally, since GBs are high flux paths as discussed earlier, increasing their density also results in a larger flux.

\begin{figure}[hbt!]
  \includegraphics[width=8cm]{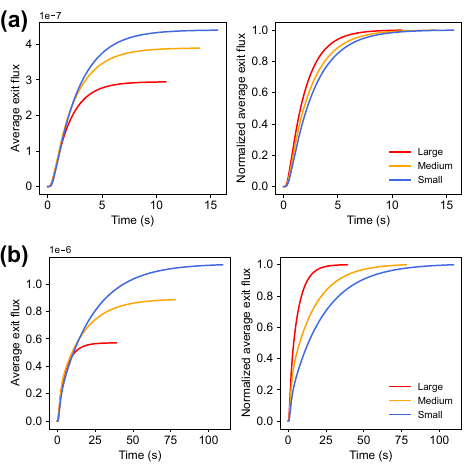}
  \centering
  \caption{Average and normalized exit flux curves setting all GBs as (a) LAGBs and (b) HAGBs.}
  \label{Fig: FluxAverage_H_LAGB}
\end{figure}

Yet, since the ratio of HAGBs/LAGBs changes with changing the grain size, two sets of simulations were performed setting all GBs as LAGBs and HAGBs in \textbf{Figure} \ref{Fig: FluxAverage_H_LAGB} a and b respectively in order to limit the discussion to the effect of grain size and rule out the effect of the GB misorientation. A similar trend can be observed in both cases, that is an increased break-through time and increased flux with decreasing grain size. The difference is that larger trap-binding energy (HAGBs) will have larger break-through time and flux compared to the smaller trap binding energy (LAGBs), which is in agreement with our previous discussions. 

\FloatBarrier

\section{Conclusions}

In this work, the effect of grain boundary misorientation on hydrogen transport was investigated using a novel phase-field based diffusion and trapping model. The model was applied to three RVEs with different grain sizes and were in good qualitative agreement with the experimental results reported in \cite{Koyama2017}. The main conclusions can be summarized as follows:

\begin{itemize}
  \item The model could capture the higher hydrogen enrichment of high-angle grain boundaries compared to the low-angle grain boundaries. Additionally, the character of the triple junctions could also be modeled.
  \item For the uptake simulations, decreasing the grain size lead to larger uptake of hydrogen, as well as longer time for saturation. This is due to the larger trap density with decreasing the grain size. 
  \item The permeation simulations showed that grain boundaries are high hydrogen flux paths, where the flux magnitude generally increases with increasing the grain boundary trap-binding energy. Furthermore, it also depends on the alignment of the grain boundary with respect to the external occupancy gradient and the network connectivity of the grain boundaries. 
  \item The early stage permeation simulations showed that high-angle grain boundaries are faster paths than low-angle grain boundaries. 
  \item Decreasing the grain size resulted in an increase in the break-through time and the steady-state exit flux.
  \item This work paves the way for more quantitative approach for designing grain boundary engineered alloys that can mitigate hydrogen embrittlement.
\end{itemize}

\textbf{Supporting Information} \par 
Supporting Information is available from the Wiley Online Library or from the author.

\textbf{Acknowledgements} \par 
AH is supported by Research Foundation -- Flanders Marie Skłodowska-Curie actions - Seal of Excellence (MSCA SoE FWO), grant number 12ZZN23N. BK acknowledges the National Research Foundation of Korea (NRF) by grant NRF-2018R1A6A1A0302552617 of the Priority Research Program under the Ministry of Education, and in part by grant NRF-2021R1A2C1004540 under the Ministry of Science and ICT.

\bibliographystyle{unsrtnat}
\bibliography{refs_Misori}

\end{document}